\documentclass{article} 
\usepackage{iclr2027_conference,times}

\usepackage{amsmath,amsfonts,bm}

\def\eqref#1{equation~\ref{#1}}

\def\1{\bm{1}}

\DeclareMathAlphabet{\mathsfit}{\encodingdefault}{\sfdefault}{m}{sl}
\SetMathAlphabet{\mathsfit}{bold}{\encodingdefault}{\sfdefault}{bx}{n}

\usepackage[utf8]{inputenc} 
\usepackage[T1]{fontenc}    
\usepackage{xurl} 

\usepackage{wrapfig}

\usepackage{hyperref}
\usepackage{url}
\usepackage{graphicx}
\newcommand{\sysname}{\textsc{AgentBug-Smith}}
\newcommand{\ourbench}{\textsc{Live-Harness-Bench}}

\newcommand{\openclaw}{OpenClaw}

\newcommand{\swefactory}{SWE-Factory}

\newcommand{\agentissuebench}{AgentIssue-Bench}

\newcommand{\gpt}{GPT-4.1-mini}
\newcommand{\kimi}{Kimi-k2.5}
\newcommand{\deepseek}{DeepSeek-v3.2}

\newcommand{\swebenchlive}{SWE-bench-Live}

\newcommand{\rawaiapi}{Vanilla LLM}

\newcommand{\minisweagent}{mini-SWE-agent}
\newcommand{\autocoderover}{AutoCodeRover}

\newcommand{\openhands}{OpenHands}
 
\usepackage{fontawesome5}

\newcommand{\parabf}[1]{\noindent\textbf{#1}}
 
\newcommand{\githubicon}{\raisebox{-0.12ex}{\scalebox{0.82}{\faGithub}}}
\newcommand{\github}{\mbox{\githubicon\,GitHub}}

\newcommand{\hficon}{\raisebox{-0.28ex}{\includegraphics[height=1.05em]{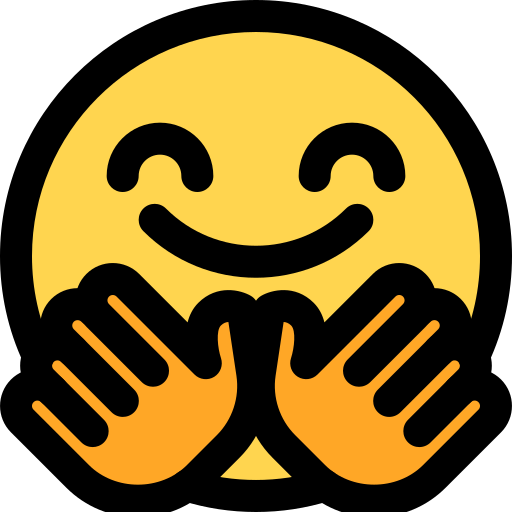}}}
\newcommand{\codeicon}{\raisebox{-0.28ex}{\includegraphics[height=1.05em]{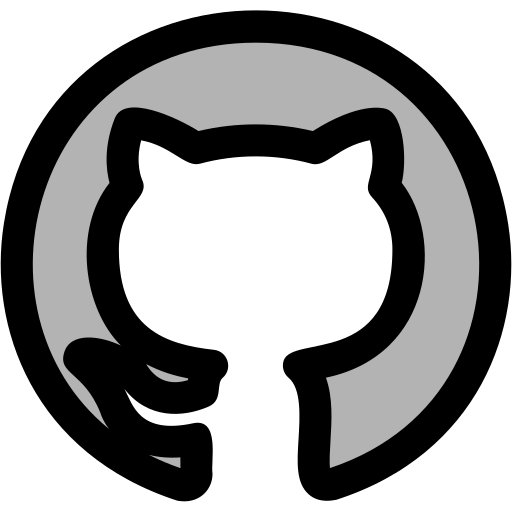}}}
\newcommand{\codeurl}{https://github.com/EaminC/AgentBug-Smith}
\newcommand{\dataurl}{https://huggingface.co/buckets/EaminChan/live-harness-bench}
\newcommand{\coderelease}{\href{\codeurl}{\codeicon~\sysname}}
\newcommand{\datarelease}{\href{\dataurl}{\hficon~\ourbench}}

\newcommand{\noopsort}[1]{}

\usepackage{adjustbox}
\usepackage{makecell}

\newcommand{\distance}{4pt}
\usepackage{multirow}
\usepackage{array}
\usepackage{makecell}
\usepackage{adjustbox}
\usepackage{algorithm}
\usepackage{algorithmic}
\usepackage{booktabs}
\usepackage{tabularx}
\usepackage{float}
\usepackage{enumitem}
\usepackage{needspace}
\usepackage{listings}
\usepackage{textcomp}
\usepackage{xcolor}
\usepackage{colortbl}
\usepackage[skins,breakable,listings]{tcolorbox}
\usetikzlibrary{tikzmark}
\usepackage{caption}
\newcolumntype{Y}{>{\raggedright\arraybackslash}X}
\definecolor{edbg}{HTML}{1E1E1E}
\definecolor{edtitle}{HTML}{2B2B2B}
\definecolor{edkey}{HTML}{E06C75}
\definecolor{edstr}{HTML}{E5C07B}
\definecolor{ednum}{HTML}{98C379}
\definecolor{edkw}{HTML}{56B6C2}
\definecolor{edpunct}{HTML}{D4D4D4}
\definecolor{edln}{HTML}{636D83}
\definecolor{edadd}{HTML}{7EE787}
\definecolor{eddel}{HTML}{FFA198}

\newtcolorbox{jsoneditor}[1]{%
  enhanced, breakable,
  colback=edbg,
  colframe=black!82,
  colbacktitle=edtitle,
  coltitle=white!92,
  boxrule=0.45pt,
  arc=2pt,
  left=5pt, right=6pt, top=5pt, bottom=5pt,
  toptitle=2.5pt, bottomtitle=2.5pt,
  fonttitle=\small\ttfamily,
  title={%
    \tikz[baseline=-0.55ex]{
      \fill[red!72] (0,0) circle (1.55pt);
      \fill[yellow!85!orange] (6.6pt,0) circle (1.55pt);
      \fill[green!62] (13.2pt,0) circle (1.55pt);
    }\hspace{7pt}#1%
  },
  before upper={\color{edpunct}\raggedright},
}
\definecolor{pyblue}{HTML}{3776AB}
\definecolor{dockerblue}{HTML}{2496ED}
\definecolor{jsongold}{HTML}{CBCB41}
\definecolor{txtgray}{HTML}{6E6E6E}
\definecolor{loggreen}{HTML}{388A34}
\definecolor{edside}{HTML}{F3F3F3}
\definecolor{edsel}{HTML}{E4E6F1}
\definecolor{edfn}{HTML}{795E26}
\definecolor{edmd}{HTML}{AF00DB}
\definecolor{edop}{HTML}{0451A5}
\definecolor{edcmt}{HTML}{008000}
\definecolor{mdblue}{HTML}{007ACC}
\definecolor{folderyellow}{HTML}{C09553}
\definecolor{idebg}{HTML}{FFFFFF}
\definecolor{idetitle}{HTML}{DDDDDD}
\definecolor{idetext}{HTML}{333333}
\definecolor{idemuted}{HTML}{6E6E6E}

\newtcolorbox{ideeditor}[1]{%
  enhanced,
  colback=white,
  colframe=black!28,
  colbacktitle=idetitle,
  coltitle=black!78,
  boxrule=0.45pt,
  arc=2pt,
  left=0pt, right=0pt, top=0pt, bottom=0pt,
  toptitle=2.5pt, bottomtitle=2.5pt,
  fonttitle=\small\ttfamily,
  sidebyside,
  sidebyside align=top,
  lefthand width=0.27\textwidth,
  sidebyside gap=0pt,
  segmentation style={black!18, solid, line width=0.8pt},
  underlay={%
    \begin{tcbclipinterior}%
      \fill[edside] (interior.north west) rectangle (segmentation.south);%
    \end{tcbclipinterior}%
  },
  before upper={\color{idetext}},
  before lower={\color{idetext}},
  title={%
    \tikz[baseline=-0.55ex]{
      \fill[red!72] (0,0) circle (1.55pt);
      \fill[yellow!85!orange] (6.6pt,0) circle (1.55pt);
      \fill[green!62] (13.2pt,0) circle (1.55pt);
    }\hspace{7pt}#1%
  },
}
\newtcolorbox{partbox}[2]{%
  enhanced,
  colback=#1!6,
  colframe=#1!28,
  colbacktitle=#1!12,
  coltitle=black,
  boxrule=0.45pt,
  arc=1.2pt,
  left=5pt, right=5pt, top=4pt, bottom=4pt,
  fonttitle=\small\bfseries,
  toptitle=1pt, bottomtitle=1.5pt,
  before upper={\raggedright},
  title={#2},
}
\newtcolorbox{specbox}[1]{%
  enhanced, breakable,
  colback=#1!6,
  colframe=#1!28,
  boxrule=0.45pt,
  arc=1.2pt,
  left=6pt, right=6pt, top=5pt, bottom=5pt,
  before upper={\raggedright\small},
}

\tcbset{
  caseinner/.style={
    enhanced,
    boxrule=0.55pt,
    arc=2.2pt,
    left=5pt, right=5pt, top=4pt, bottom=4pt,
    toptitle=2.5pt, bottomtitle=2.5pt,
    fonttitle=\small\bfseries,
    before skip=3.5pt,
    after skip=3.5pt,
  }
}
\newtcolorbox{caseouter}{%
  enhanced, breakable,
  colback=white,
  colframe=black!25,
  boxrule=0.65pt,
  arc=3pt,
  left=4.5pt, right=4.5pt, top=4pt, bottom=4pt,
  before skip=6pt, after skip=8pt,
}
\newtcolorbox{caseissue}{%
  caseinner,
  colframe=black!32,
  colback=white,
  colbacktitle=black!7,
  coltitle=black!62,
  title={Issue},
  before upper={\raggedright\small},
}
\newtcolorbox{casefail}{%
  caseinner,
  colframe=red!48,
  colback=white,
  colbacktitle=red!10,
  coltitle=red!55!black,
  title={Without the skill},
  before upper={\raggedright\small},
}
\newtcolorbox{casegen}{%
  caseinner,
  colframe=red!48,
  colback=white,
  colbacktitle=red!10,
  coltitle=red!55!black,
  title={Generated patch},
  before upper={\raggedright\small},
}
\newtcolorbox{casegold}{%
  caseinner,
  colframe=green!45!black!35,
  colback=white,
  colbacktitle=green!12,
  coltitle=green!35!black,
  title={Golden patch},
  before upper={\raggedright\small},
}
\newcommand{\casefile}[1]{{\small\color{black!50}\textit{#1}}\par\vspace{2.5pt}}
\newcommand{\difflinem}[1]{%
  \par\nobreak\noindent
  \colorbox{red!14}{%
    \parbox{\dimexpr\linewidth-2\fboxsep\relax}{%
      \raggedright\ttfamily\small\textcolor{red!70!black}{\mbox{$-$} #1}}}%
  \par}
\newcommand{\difflinep}[1]{%
  \par\nobreak\noindent
  \colorbox{green!14}{%
    \parbox{\dimexpr\linewidth-2\fboxsep\relax}{%
      \raggedright\ttfamily\small\textcolor{green!40!black}{\mbox{$+$} #1}}}%
  \par}

\title{\sysname\,\raisebox{-0.18em}{\includegraphics[height=1.05em]{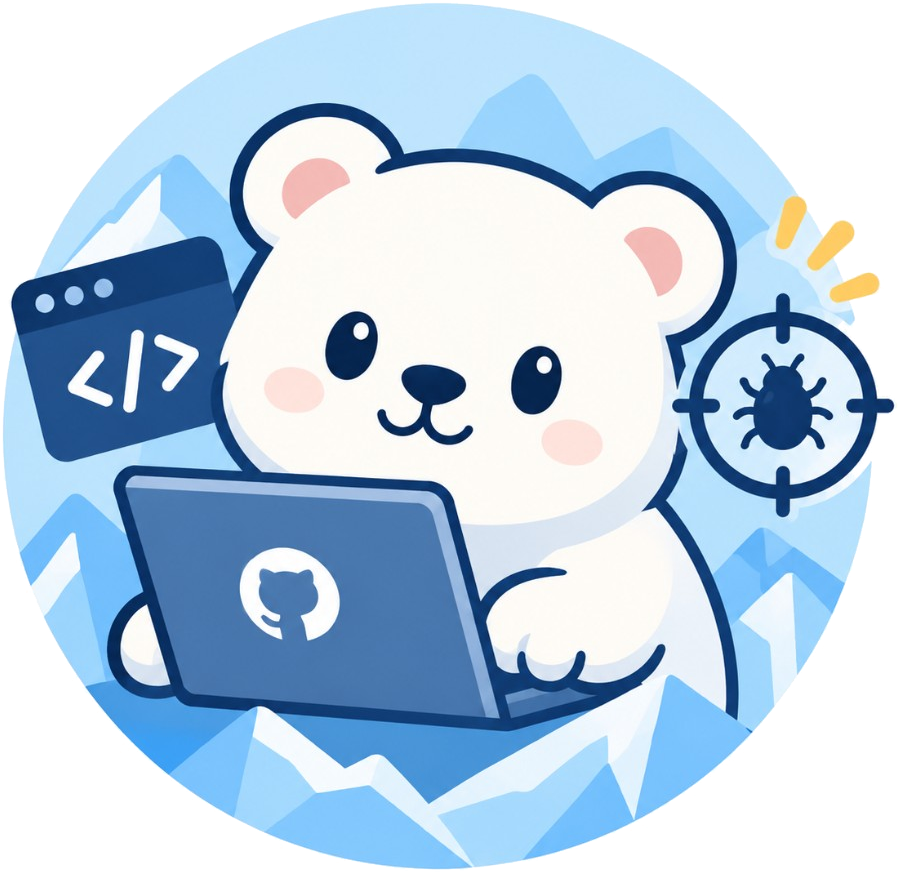}}:\\[0.12em]
Automatically Reproducing Real-World\\
Harness Bugs in Agentic Systems}

\author{
Yiming Cheng \\
The University of Chicago \\
\texttt{eaminchan@uchicago.edu}
\And
Alfin Wijaya Rahardja \\
Fudan University \\
\texttt{24212010055@m.fudan.edu.cn}
\And
Mengshi Zhang \quad and \quad Zihao Chen \\
TensorBlock, Inc. \\
\texttt{\{mengshiz, zihaoc\}@tensorblock.co}
\And
Zhenpeng Chen \\
Tsinghua University \\
\texttt{zpchen@tsinghua.edu.cn}
\And
Yiling Lou\\
University of Illinois Urbana-Champaign \\
\texttt{yilingl@illinois.edu}
}

\iclrfinalcopy 
\begin{document}

\maketitle

\begin{abstract}
Agent harness bugs exhibit unique characteristics and remain challenging for state-of-the-art software agents to repair. Progress in this area is further hindered by existing benchmarks, which contain only a small and fixed number of executable harness bugs while requiring hundreds of human hours to construct. 
This work presents \sysname{}, an automated harness bug reproduction approach that continuously discovers and reproduces real-world harness bugs from open-source agentic systems. Across different backbone LLMs, \sysname{} consistently outperforms existing bug reproduction techniques designed for general software, achieving 10.67\% - 27.56\% higher success rates of reproducing harness bugs. By applying \sysname{} to open-source agentic systems in the wild, we construct \ourbench{}, a live and extensible benchmark that currently contains 200 reproducible harness bugs. We further demonstrate the utility of \ourbench{} through two downstream applications. First, we use \ourbench{} as the evaluation benchmark to systematically evaluate state-of-the-art software agents, revealing their limited capabilities in repairing real-world harness bugs. Second, we use \ourbench{} as a knowledge base of real-world harness bug fixes, from which reusable repair skills can be distilled to improve existing software agents, increasing their harness-bug repair rates by 6.32\%. Together, \sysname{} and \ourbench{} establish a scalable foundation for continuously evaluating and improving software agents on harness bug repair, turning real-world agent failures into executable evaluation instances and reusable knowledge for harness improvement, thus contributing to the ultimate goal of recursively self-improving agents. 
\end{abstract}
\begin{center}
\vspace{-0.6em}
\coderelease\hspace{1.6em}\datarelease
\end{center}
\section{Introduction}
Large Language Model (LLM) agents are rapidly emerging as a new software paradigm and have been increasingly adopted across diverse domains~\citep{swe-agent,  DBLP:conf/iros/KannanVM24,DBLP:journals/corr/abs-2311-07226, DBLP:journals/corr/abs-2502-18864,corrabs250519562}.  Agentic systems are commonly composed of backbone LLMs and their surrounding \emph{harnesses}, where the harness serves as a software infrastructure that works with LLMs to jointly shape the agent behavior. Modern agents involve increasingly complex and large-scale harness implementations (e.g., 200K lines of code in \openclaw{}~\citep{openclaw}) to support diverse functionalities such as context management, security guardrails, orchestration, and tool integration. Accordingly, there have been growing research efforts~\citep{li2026agentharness} on harness engineering, demonstrating that improving the harness alone can substantially enhance agent effectiveness, even when the underlying backbone LLM remains unchanged. More recently, advances in self-evolving harnesses~\citep{lee2026meta} and recursive self-improvement~\citep{liu2026path} have further elevated the harness from a static supporting layer to an evolving component of agentic systems, underscoring the critical role of harness quality in determining the agent capabilities.

The growing complexity of agent harnesses inevitably makes \emph{harness implementation bugs} an important source of failures in agentic systems. Recent studies~\citep{neuripsRahardja25, zhang2026understanding, chen2026understanding} have revealed that modern agent applications and frameworks suffer from diverse implementation bugs in harnesses. Automatically detecting and repairing such bugs is therefore becoming increasingly critical for building reliable agentic systems. However, prior work~\citep{neuripsRahardja25} has shown that existing software agents~\citep{swe-agent, xia2024agentless, auto-code-rover} achieve substantially lower success rates in repairing harness bugs than general software bugs (e.g., 4.67\% versus 40.67\%). This difficulty stems in part from the distinct characteristics of harness bugs, which arise from agent-specific structures and behaviors, such as complex workflow orchestration and interactions with external resources (e.g., tools, services, and model providers). These characteristics introduce failure modes and repair challenges that are uncommon in traditional software systems. Taken together, automated harness bug repair represents an emerging and distinct challenge in improving agent reliability. 

However, we still lack comprehensive benchmarks of harness bugs. Most benchmarks of agent failures~\citep{MAST, zhu2025llm} focus on failures stemming from backbone LLMs rather than harness implementation bugs. While there have been increasing effort to investigate harness implementation bugs~\citep{zhang2026understanding, chen2026understanding}, existing bug collections lack executable environments and artifacts for reproducing the agent failures. To date, the only executable benchmark of reproducible harness bugs,  \agentissuebench{}~\citep{neuripsRahardja25}, contains only a small and fixed set of 43 harness bugs, which cannot support comprehensive or rigorous evaluation due to the limited number of bugs and the potential risk of contamination from exposure to model training data. In addition, manually reproducing harness bugs is labor-intensive and time-consuming, e.g., 150 manual hours were spent in reproducing a small number of  harness bugs~\citep{neuripsRahardja25}.

While recent approaches such as \swefactory{}~\citep{guo2025swefactory} and \swebenchlive{}~\citep{zhang2025swebenchgoeslive} have automated the reproduction of bugs reported in GitHub Issues, they are primarily designed for general software systems. Harness bugs in agentic systems pose unique reproduction challenges, as agent harnesses extensively interact with external resources, including LLM providers, tools, network services, and dynamically changing environments, while their failures often manifest only under specific execution states and interaction contexts. These characteristics fundamentally distinguish harness bugs from conventional software bugs and pose substantial challenges to directly applying existing automated bug reproduction techniques to agentic systems.

\parabf{Technique.} This work proposes \sysname{}, an automated harness bug reproduction approach that continuously discovers and reproduces real-world harness bugs from open-source agents. \sysname{} is built as a multi-agent system to streamline reproduction pipeline without any manual intervention, facilitating three phases including (i) harness bug related issues identification, (ii) agent execution environment construction, and (iii) failure-reproducing harness test generation.

\parabf{Evaluation.}  We perform extensive evaluation to show the effectiveness of \sysname{}. First, based on our manual checking,  \sysname{} achieves a high accuracy of automatically identifying agent repositories (i.e., 95\% accuracy) and harness bug related issues (i.e., 92\%). Second, \sysname{} consistently achieves higher success rate of reproducing harness bugs compared state-of-the-art bug reproduction techniques that are designed for general software (i.e., \swefactory{} and \swebenchlive{}), with 10.67\% - 27.56\% percentage-point  improvements across all studied backbone LLMs. Third, our ablation analysis further confirms the effectiveness of both environment and test generation components in \sysname{} compared to baselines. 

\parabf{Benchmark and Downstream Application.}  Continuously applying \sysname{} to open-source agentic systems in the wild can enabling scalable construction of a benchmark of executable harness bugs, \ourbench{}. Its current release contains 200 reproducible harness bugs collected from real-world agent repositories,  covering diverse harness components and spanning over wide time ranges. 
We further demonstrate the usage of \ourbench{} in two downstream applications.  First, we use \ourbench{} as the evaluation benchmark to systematically evaluate state-of-the-art software agents, revealing their limited capabilities in repairing real-world harness bugs. Second, we use \ourbench{} as a knowledge base of real-world harness bug fixes, from which reusable repair skills can be distilled to improve existing software agents, increasing their harness-bug repair success rate by 6.32\%. Together, \sysname{} and \ourbench{} establish a scalable foundation for continuously evaluating and improving software agents on harness bug repair, turning real-world agent failures into executable evaluation instances and reusable knowledge for harness  improvement, thereby contributing to the ultimate goal of recursively self-improving agents.

\section{Related Work}

\parabf{Failures in Agentic Systems.}
Substantial research efforts have been devoted to analyzing and understanding failures in agentic systems. Most existing studies or benchmarks~\citep{whoandwhen, zhu2025llm, MAST, bouzenia2025understanding} focus on the runtime failures stemming from the underlying LLMs but not the failures caused by implementation bugs in the agent harness layer. Rahardja et al.~\cite{neuripsRahardja25} performed the first study to investigate the implementation bugs in agentic systems, summarizing diverse categories of bugs across different agent components and building the first executable benchmark of agent bugs,  \agentissuebench{}. More recently, there have been increasing effort in investigating the bugs in agent harnesses~\citep{zhang2026understanding, chen2026understanding, zhu2026bugs, shao2024llms, xue2025characterization}, they do not provide executable environments or artifacts to trigger the failure. To date, \agentissuebench{} is the only benchmark of executable harness bugs, but only includes a small and fixed number of bugs (i.e., 43) and is constructed with huge manual effort (i.e., 150 hours).  To date, we still lack large-scale and comprehensive benchmarks of reproducible agent harness bugs.

\parabf{Automated Bug Benchmark Construction.}
Existing techniques have automated the construction of software bug benchmarks and datasets~\citep{yang2026swe, jain2025r2e}, but primarily focus on synthesizing large numbers of  artificial bugs through mutation rather than reproducing real-world bugs. Such datasets are often used as training data for model improvement. In this work, we instead focus on constructing benchmarks of real-world bugs to enable more realistic bug distribution in practice.
While there has been prior work~\citep{zhang2025swebenchgoeslive, guo2025swefactory, badertdinov2026swe, tomassi2019bugswarm, pan2024training} that automatically construct benchmarks from real-world bugs in open-source software (e.g., mostly Github issues), they are designed for general software. In contrast, this work specifically focuses on reproducing real-world bugs in agentic systems, which pose unique reproduction challenges due to the extensive interactions between agent harnesses and external resources.

\section{\sysname}

\begin{figure}[t]
    \centering
    \includegraphics[width=0.9\linewidth]{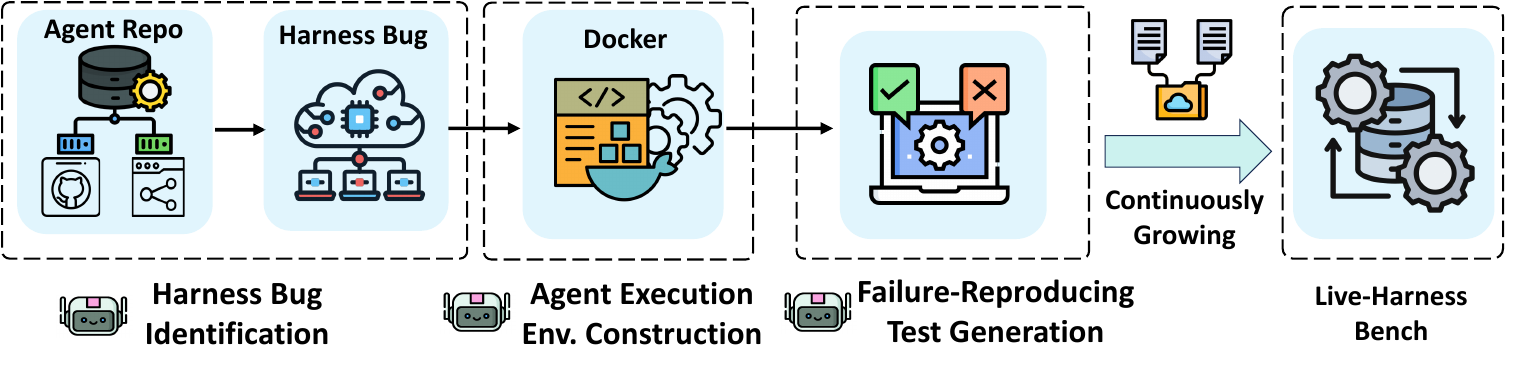}
    \caption{Overview of \sysname{}.}
    \label{fig:overview}
\end{figure}

In this section, we present \sysname, a fully automated approach that reproduces real-world harness bugs from open-source agentic systems and thus can continuously construct a growing benchmark. Following widely-adopted construction pipelines for general-software bug benchmarks~\citep{sweboard, guo2025swefactory, zhang2025swebenchgoeslive}), we use GitHub issues as the source of real-world agent failures. Figure~\ref{fig:overview} presents the overview of \sysname{}, which is built as multi-agent systems covering the following three key stages. 
(i) \emph{Harness Bug Identification}, an agentic pipeline that discovers high-quality GitHub repositories of agentic systems and mines issues related to harness bugs (Section~\ref{method:sub1});
(ii) \emph{Agent Execution Environment Construction}, an agentic pipeline that builds environment containers required for agent execution and bug reproduction  (Section~\ref{method:sub2}); 
and (iii) \emph{Failure-Reproducing Harness Test Generation}, an agentic pipeline  that synthesizes the harness test that can exactly trigger the issue-described agent failure (Section~\ref{method:sub3}).

\begin{figure}[t]
\centering
\includegraphics[width=\linewidth]{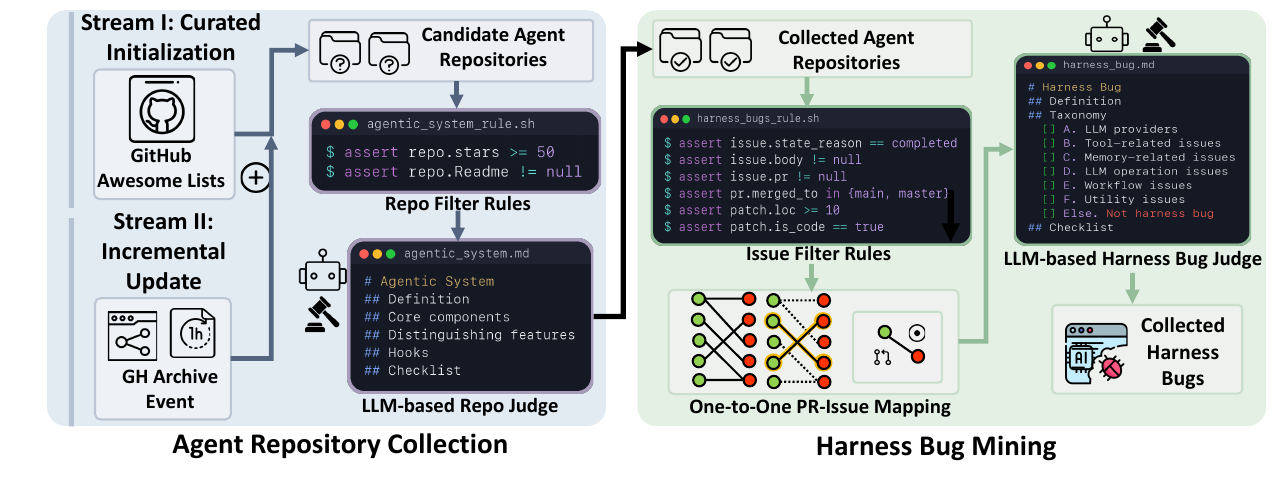}
\caption{Harness Bug Identification in \sysname{}.}
\label{fig:mining}
\end{figure}
\subsection{Harness Bug Identification}\label{method:sub1}
As the open-source agent ecosystem undergoes rapid and continuous development, static benchmarks can quickly become outdated. To capture the dynamics of this highly active community in real time, we design an automated harness bug identification pipeline to continuously discover emerging open-source agent repositories and identify GitHub issues related to agent harness bugs. Figure~\ref{fig:mining} illustrates the overall process, which consists of the following two key steps.

\parabf{Agent Repository Collection.} To maintain an up-to-date list of high-quality and widely used agent repositories, we incorporate a dual-stream repository discovery mechanism, including: (1)  \textit{Curated Initialization}, which performs a one-time, large-batch initialization of the repository pool based on community-curated awesome lists of GitHub agentic systems~\citep{kyrolabs-awesome,jenqyang-awesome,slavakurilyak-awesome,e2b-awesome,jim-schwoebel-awesome,rohitg00-awesome,georgezouq-awesome,aihubcn-awesome,foundationagents-awesome,shubhamsaboo-awesome,hyp1231-awesome,wangrongsheng-awesome,tensorchord-awesome,Awesome-LLM-Inference}; (2) \textit{Incremental Stream Update}, which continuously processes near-real-time GitHub Archive event streams in small batches to discover active and emerging agent repositories based on developer interactions. The repositories identified through these two streams are merged into a unified candidate pool, allowing the initial curated collection to be continuously expanded with newly emerging projects.

To ensure the quality of selected repositories, we apply two inclusion criteria. First, repositories must demonstrate sufficient community adoption (i.e., at least $50$ GitHub stars) and contain tests, providing basic support for subsequent environment construction and reproduction test generation. Detailed filtering rules are in Appendix~\ref{app:repo-rules}. Second, repositories must follow commonly adopted agent structures, involving components such as tools, orchestration, and provider integrations. Specifically, following common agent and harness components defined in prior work~\citep{arxivguo26,li2026agentharness,Agent4SE,meng2026agentharness,Wang_2024}, we design an LLM-based judge to analyze each candidate repository and check  whether it implements an agentic system. The detailed specification provided to the LLM judge is included in Appendix~\ref{app:agentic-spec}.

\parabf{Harness Bug Mining.} \sysname{} then mines GitHub issues related to harness bugs from each agent repository. First, to ensure the quality of selected issues, we follow a set of rules (detailed in Appendix~\ref{app:issue-rules}) requiring each issue to form a valid \textit{issue-pull request (PR) pair}, where the associated PR has been successfully merged and contains functional code changes. Moreover, as issues, PRs, and commits often exhibit many-to-many relationships, we further perform one-to-one matching (detailed in Appendix~\ref{app:one-one-match}) to identify the unique buggy and patched commits for each issue.

Second, as revealed in previous work~\citep{neuripsRahardja25}, not all GitHub issues in agent repositories are related to harness implementation bugs. In fact, a non-trivial proportion of GitHub issues involve common bugs (e.g., utility bugs) that can also occur in general software systems. Therefore, to ensure that our constructed benchmark focuses specifically on harness bugs rather than being mixed with general software bugs, we leverage LLMs to inspect each issue and its patch, retaining an issue only if its patch occurs in a harness component. In particular, we follow the common definition of agent harness components~\citep{arxivguo26,li2026agentharness,meng2026agentharness,arxivning26}, e.g., orchestration, memory, or tools, and design a harness bug specification for the LLM judge (detailed prompt in Appendix~\ref{app:harness-spec}).

After this stage, \sysname{} returns a set of candidate harness bugs, whose metadata include the issue description text, the linked PR, the buggy commit, and the patched commit.

\subsection{Agent Execution Environment Construction}\label{method:sub2}
\begin{figure}[t]
\centering
\includegraphics[width=\linewidth]{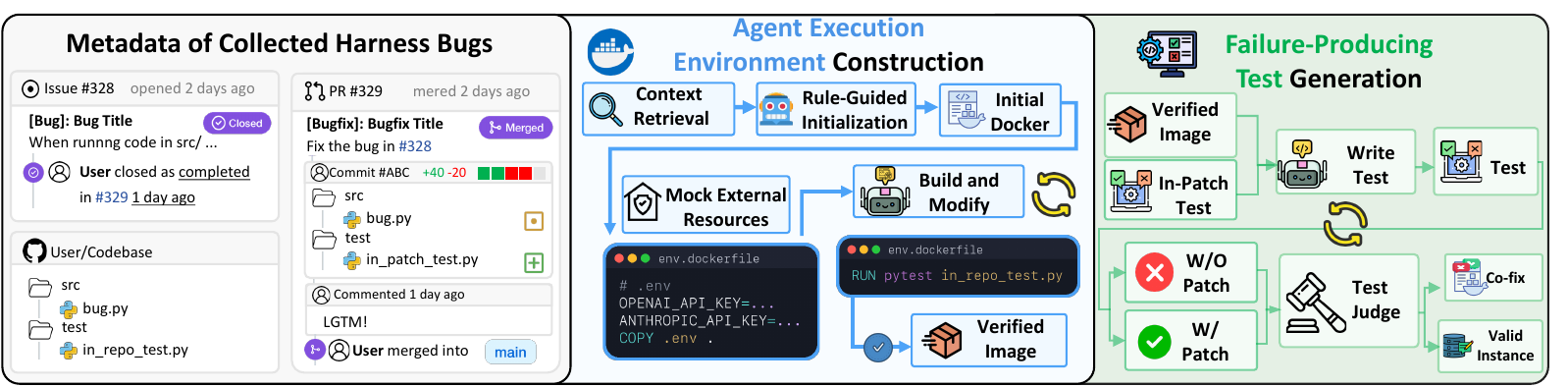}
\caption{Environment Construction and Failure-Reproducing Test Generation.}
\label{fig:system-overview}
\end{figure}

Compared to general software, agentic systems often depend heavily on external resources (e.g., model providers, tools, and external knowledge bases), making the setup of a proper execution environment a critical step in reproducing user-reported harness bugs. While existing techniques~\citep{guo2025swefactory, zhang2025swebenchgoeslive, hu2026repo2run} can generate execution environments for general software, \sysname{} introduces an environment generation pipeline that automatically synthesizes a Dockerfile ($\mathcal{D}$) for constructing an isolated execution environment specifically tailored to agent-dependent execution resources. In particular, to enforce strict isolation during evaluation, \sysname{} injects mock service environment variables at the container level, transparently rerouting external model client invocations to localized test harnesses without requiring intrusive modifications to the underlying codebase. As illustrated in Figure~\ref{fig:system-overview}, \sysname{} adopts the following steps for environment construction.

\parabf{Rule-Guided Initialization.} Rather than relying on LLMs to generate Dockerfiles from scratch, \sysname{} first performs a rule-based scanning routine that parses standard package descriptors, including dependency manifests, build configurations, and CI/CD workflows, to automatically infer runtime configurations (e.g., programming language, version constraints, package manager, installation commands, and test runner). Based on the collected context, \sysname{} generates an initial Dockerfile $\mathcal{D}_{init}$ from an official minimal image with a single LLM API invocation, thereby incurring only minimal LLM inference cost.

\parabf{Iterative Environment Refinement.} The initial Dockerfile $\mathcal{D}_{init}$ can be imperfect (e.g., omitting a system library, specifying an incompatible package version, or placing the repository on an incorrect import path). \sysname{} therefore incorporates a tool-using agent to iteratively refine $\mathcal{D}$ based on build error messages. If the build succeeds, \sysname{} further samples a set of existing tests (e.g., 20) and executes them as an environment smoke test. We intentionally use a lightweight sample rather than the full test suite, as the goal at this stage is to validate basic environment executability rather than repository correctness. Running the full test suite would not only incur substantial overhead during iterative refinement, but could also introduce false-positive environment failures from buggy, flaky, or external-resource-dependent tests. Accordingly, assertion failures are excluded from the diagnostic feedback, allowing the agent to focus on environment-related failures such as collection, import, and dependency errors. The refinement loop terminates when the environment is successfully constructed or the maximum number of attempts is reached. In particular, the environment is considered successfully constructed once the image builds successfully and the tests can pass the execution.

\subsection{Failure-Reproducing Harness Test Generation}\label{method:sub3}
A failure-reproducing test is a test that fails on the buggy commit with the same failure  reported in the corresponding issue description, while passing on the patched commit. In practice, some bug-fixing PRs contain both the patch and an \textit{in-patch test} (IPT), where the IPT is submitted alongside the patch to validate its correctness and can therefore naturally serve as a failure-reproducing test. However, IPTs are not always available: according to our statistics, less than half (47.11\%) of the issues contain IPTs. Restricting reproduction only to issues with IPTs would therefore substantially limit the scope of reproducible bugs and bias the resulting benchmark toward harness bugs for which developers explicitly provided tests. To broaden the coverage of reproducible harness bugs to issues both with and without IPTs, \sysname{} introduces an agentic workflow that automatically synthesizes failure-reproducing tests for a given harness bug, even when an IPT is absent.

The LLM is provided with the corresponding context (e.g., the buggy code repository, issue description, and developer-submitted patch) to generate an initial test $\mathcal{T}$ that triggers the failure described in the issue. To ensure faithful reproduction, the generated test must import and exercise the actual buggy implementation rather than replacing the target functionality with mocks. Mocking is permitted only at model-client or network boundaries to isolate test execution from nondeterministic external networks and third-party model APIs, which is consistent with common testing practices in agentic systems~\citep{hasan2026empirical}. In contrast, functions under test cannot be mocked, ensuring that the reproduced failure originates from the actual buggy implementation.

\parabf{Joint Optimization of Test and Environment.}
A generated test is considered failure-reproducing only if (i) it fails on the buggy commit with the same failure reported in the issue description and (ii) it passes on the patched commit. If either condition is not satisfied, \sysname{} iteratively refines the test $\mathcal{T}$ based on its execution outcomes until both conditions are satisfied or the maximum number of iterations is reached.
However, optimizing the test $\mathcal{T}$ alone can sometimes be insufficient, as unsuccessful reproduction may stem from misalignment between the generated test $\mathcal{T}$ and the environment Dockerfile $\mathcal{D}$. For example, a generated test may exercise an incorrect assertion while the constructed environment simultaneously imports an installed package instead of the checked-out source. To address such cases, beyond test-only optimization, \sysname{} optionally performs joint optimization to co-refine the  test $\mathcal{T}$ and the environment Dockerfile $\mathcal{D}$. Appendix~\ref{app:cofix-case} presents an example where such joint optimization successfully brings the test into a fail-to-pass state (i.e., failing on the buggy commit while passing on the patched commit), whereas test-only optimization cannot.

\section{Evaluation}

\subsection{Effectiveness of Harness Bug Identification}
This section first evaluates the effectiveness of the first component (i.e., harness bug identification) in \sysname{}, as it serves as the foundation for all the subsequent components. 

\parabf{Settings.} For continuous harness bug identification, we primarily use cost-effective models (i.e., \gpt{}) in \sysname{}, as this stage continuously processes a large volume of issues and thus requires cost-effective model inference for scalable deployment. As there is no existing technique specifically designed for harness bug identification, we compare our approach against vanilla LLM invocation as the baseline. To evaluate identification accuracy, we randomly sample 132 out of 225 GitHub issues (corresponding to a 95\% confidence level and a 5\% margin of error) identified as harness-bug-related by our approach and the vanilla baseline. Two annotators then independently label whether each selected repository is an agent repository and whether each selected issue is related to a harness bug. The resulting Cohen's kappa is 0.88, indicating high inter-annotator agreement~\citep{landis1977measurement}. In addition, given the inherent randomness of LLM inference, we evaluate the stability of our approach by independently classifying each identified issue five times and reporting the set agreement $S_{\mathrm{set}}$ and normalized entropy $S_{\mathrm{entropy}}$ (formulas in Appendix~\ref{app:stability}).

\begin{table}[!htbp]
\centering
\small
\caption{Accuracy and Stability of Harness Bug Identification.}
\label{tab:identification}
\begin{adjustbox}{width=0.6\columnwidth}
\begin{tabular}{c|rr|rr}
\toprule
\textbf{Method} & \multicolumn{2}{c|}{\textbf{Stability}} & \multicolumn{2}{c}{\textbf{Accuracy}} \\
\cline{2-3} \cline{4-5}
 & $s_{\mathrm{set}}$ & $s_{\mathrm{entropy}}$ &  Repo. Acc. & Issue Acc. \\ \midrule 
\rawaiapi & 0.72 & 0.89 & 0.35 & 0.46  \\
\sysname & 0.92 & 0.94 & 0.95 & 0.92  \\
\bottomrule
\end{tabular}
\end{adjustbox}
\end{table}

\parabf{Results.} As shown in Table~\ref{tab:identification}, \sysname{} achieves high accuracy in identifying both agent repositories (i.e., 95\% repository-level accuracy) and harness bug related issues (i.e., 92\% issue-level accuracy). Moreover, our approach demonstrates higher stability across repeated executions than vanilla LLM invocation, further supporting the reliability of the issues identified by \sysname{}. Overall, these results demonstrate that \sysname{} can accurately and consistently identify harness bug related issues in practice.

\subsection{Reproduction Success Rate}
This section evaluates the bug reproduction success rate of \sysname{} and compares it with state-of-the-art bug  reproduction techniques that are designed for general software systems.

\parabf{Baselines and Settings.} We include two state-of-the-art bug reproduction techniques, \swefactory{}~\citep{guo2025swefactory} and \swebenchlive{}~\citep{zhang2025swebenchgoeslive}, as our baselines and directly adopt their released implementations. Following previous work~\citep{guo2025swefactory}, we evaluate all studied techniques using three different and widely used backbone LLMs, 
\begin{wrapfigure}{r}{0.43\textwidth}
    \centering
    \captionof{table}{Reproduction Success Rate.}\label{tab:successrate}
        \vspace{0pt}
        \centering
        \small
        \resizebox{\linewidth}{!}{%
        \begin{tabular}{l|rr}
        \toprule
        \textbf{Method} & \textbf{Success rate} & \textbf{Avg. \$} \\
        \midrule \hline
        \multicolumn{3}{c}{\cellcolor{gray!25}\textbf{\gpt{}}} \\ \hline
        \swefactory{} & 21/225 (9.33\%) & 0.09 \\
        \swebenchlive{} & 6/225 (2.67\%) & 0.41 \\
        \sysname{} & 45/225 (20.00\%) & 0.56 \\
        \hline
        \multicolumn{3}{c}{\cellcolor{gray!25}\textbf{\kimi{}}} \\ \hline
        \swefactory{} & 19/225 (8.44\%) & 0.37 \\
        \swebenchlive{} & 6/225 (2.67\%) & 0.66 \\
        \sysname{} & 46/225 (20.44\%) & 0.91 \\
        \hline
        \multicolumn{3}{c}{\cellcolor{gray!25}\textbf{\deepseek{}}} \\ \hline
        \swefactory{} & 31/225 (13.78\%) & 0.31 \\
        \swebenchlive{} & 1/225 (0.44\%) & 1.05 \\
        \sysname{} & 63/225 (28.00\%) & 2.44 \\
        \bottomrule
        \end{tabular}%
        }
    \vspace{8pt}
    \includegraphics[width=0.85\linewidth]{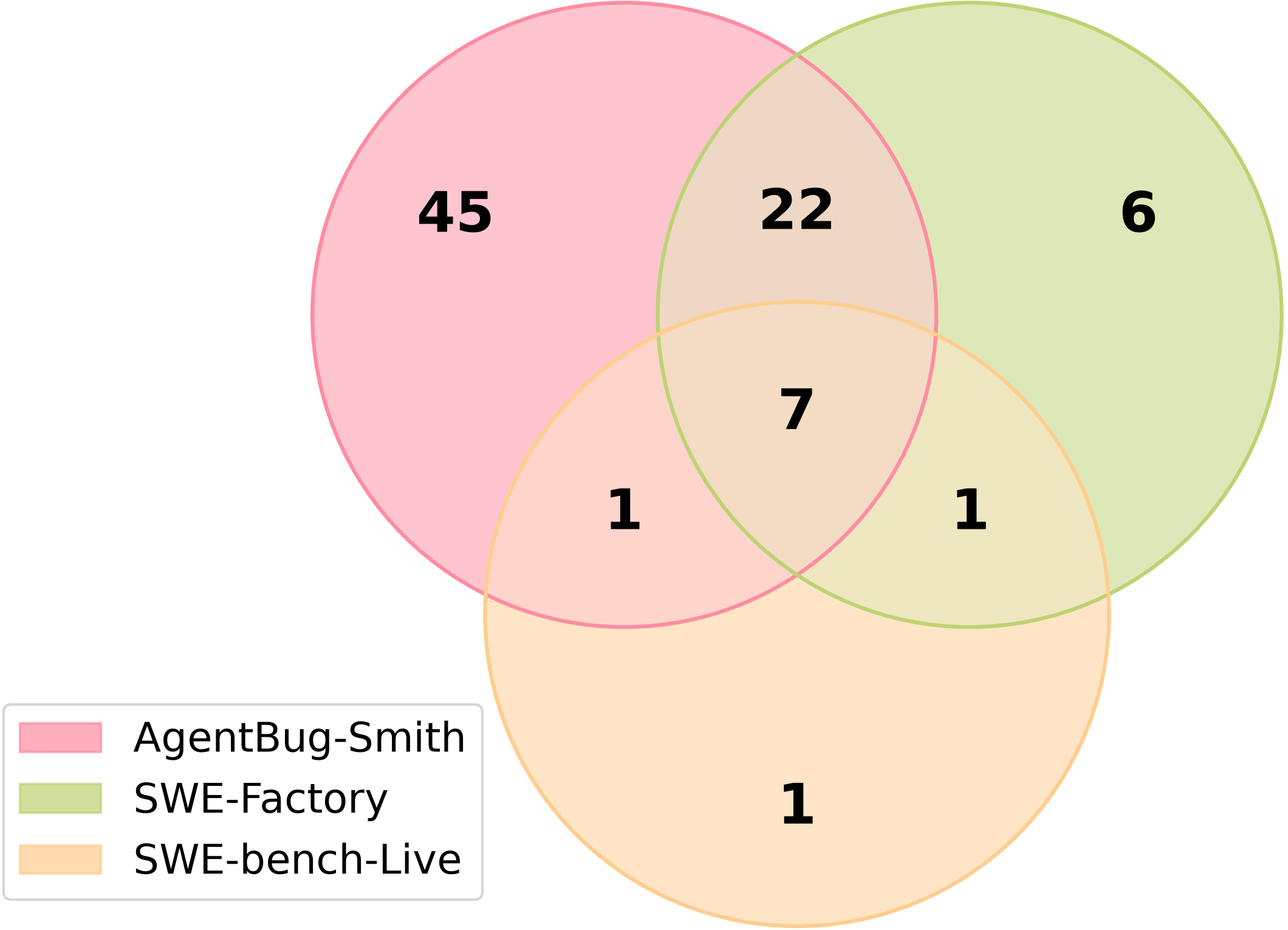}
     \vspace{2pt}
    \captionof{figure}{Overlapping Analysis.}
    \label{fig:venn}
\end{wrapfigure}
including \gpt{}~\citep{GPT4.1mini}, \kimi{}~\citep{kimik2.5}, and \deepseek{}~\citep{deepseekv3.2}, considering the cost-effectiveness required for continuous application to a large volume of GitHub issues. 
Since \swefactory{} and \swebenchlive{} are designed to reproduce general software bugs from GitHub issues, they do not include components for identifying harness bug related issues. Therefore, for a fair comparison, we apply all studied approaches to the same set of 225 harness bug-related issues identified by the first component of \sysname{} in the previous section.

\parabf{Results.} As shown in Table~\ref{tab:successrate}, despite the inherent difficulty of reproducing real-world harness bugs, \sysname{} successfully reproduces up to 28.00\% of the harness bugs, substantially outperforming existing baselines (e.g., 10.67\%-27.56\% percentage-point improvements) at acceptably higher costs (e.g., 0.56\$-2.44\$ per issue).  This result is particularly encouraging given that the best baseline can only achieve up to 13.78\% reproduction success rate. Notably, such improvements are consistent across all backbone LLMs, demonstrating the generality of \sysname{}. We further analyze the overlap among the issues reproduced by all techniques (across all backbone LLMs) in Figure~\ref{fig:venn}. Overall, \sysname{} successfully reproduces 75 bugs, including 45 unique bugs that cannot be reproduced by either baseline. These results further demonstrate the unique effectiveness of \sysname{} and highlight the necessity of developing agent-oriented bug reproduction techniques. Furthermore, to validate automatically generated tests truly reproduce the reported failure rather than merely satisfying fail-to-pass, we also perform manual inspection of all the 75 successfully-reproduced bugs, showing 74 (98.67\%) of them are valid. Nevertheless, to better understand the limitations of \sysname{}, Appendix~\ref{app:failure-modes} further performs failure analysis of  unsuccessful reproduction.

\subsection{Ablation Analysis}
This section performs the ablation analysis to investigate the effectiveness of the remaining two components (e.g., environment construction and test generation) in \sysname{}. 

\parabf{Effectiveness of Environment Construction.} In Table~\ref{tab:ablation}, the column ``Env. Build'' compares the effectiveness of environment construction in all studied techniques. In particular, \sysname{} substantially outperforms both baselines across all backbone models, i.e., 10.67\% - 35.11\% percentage-point improvements in success rate of environment construction. These results suggest that effectively constructing execution environments for agentic systems requires explicitly accounting for their agent-specific dependencies and execution characteristics. 

\begin{table}[!htbp]
\centering
\small
\caption{Effectiveness of Environment Construction and Test Generation.}
\label{tab:ablation}
\begin{adjustbox}{width=0.7\columnwidth}
\begin{tabular}{l|c|c|c|c}
        \toprule
        \textbf{Method} & \textbf{Env. Build} & \textbf{\#Test Gen.} & \textbf{w/ IPT} & \textbf{wo/ IPT} \\
        \midrule
        \hline
        \multicolumn{5}{c}{\cellcolor{gray!25}\textbf{\gpt{}}} \\ \hline
        \swefactory{} & 36/225 (16.00\%) & 21 & 21 & 0 \\
        \swebenchlive{} & 33/225 (14.67\%) & 6 & 6 & 0 \\
        \sysname{} & 84/225 (37.33\%) & 45 & 24 & 21 \\
        \hline
        \multicolumn{5}{c}{\cellcolor{gray!25}\textbf{\kimi{}}} \\ \hline
        \swefactory{} & 27/225 (12.00\%) & 19 & 19 & 0 \\
        \swebenchlive{} & 29/225 (12.89\%) & 6 & 6 & 0  \\
        \sysname{} & 53/225 (23.56\%)  & 46 & 20 & 26 \\
        \hline
        \multicolumn{5}{c}{\cellcolor{gray!25}\textbf{\deepseek{}}} \\ \hline
        
        \swefactory{} & 44/225 (19.56\%)  & 31 & 31 & 0 \\
        \swebenchlive{} & 14/225 (6.22\%) & 1 & 1 & 0\\
        \sysname{} & 93/225 (41.33\%) & 63 & 34 & 29 \\
        \bottomrule
        \end{tabular}
\end{adjustbox}
\end{table}

\parabf{Effectiveness of Test Generation.} In Table~\ref{tab:ablation}, the ``\#Test Gen.'' column reports the overall number of issues that can be successfully reproduced with failure-reproducing tests, while the ``w/ IPT'' and ``w/o IPT'' columns report the numbers of successfully reproduced issues that originally come with and without in-patch tests, respectively. Notably, \sysname{} is capable of generating failure-reproducing tests for issues without any in-patch tests, whereas both baselines fail to reproduce any such issues when no in-patch tests are available as references. This gap demonstrates the effectiveness of the failure-reproducing test generation component, which enables \sysname{} to reproduce a broader scope of harness bugs.

\subsection{\ourbench{}: Statistics and Downstream Application}
\begin{wrapfigure}{r}{0.43\textwidth}
    \centering
    \captionof{table}{Benchmark Statistics.}
    \centering
\small
\resizebox{0.83\linewidth}{!}{%
\begin{tabular}{l | l | c c}
  \toprule
  \textbf{Level} & \multicolumn{1}{c|}{\textbf{\#Item}} & \textbf{Average} & \textbf{Median} \\
  \midrule
  \multirow{2}{*}[0pt]{Repo.} 
    & LoC & 123k & 83k \\
    & Files & 479 & 343 \\
  \midrule
  \multirow{3}{*}[0pt]{Patch} 
    & Files & 3.8 & 3.0 \\
    & Hunks& 19.8 & 12.0 \\
    & Lines & 202.9 & 102.5 \\
  \bottomrule
\end{tabular}
}
\vspace{1ex}
    \label{tab:stats-bench}
    \par\vspace{2pt}
    \includegraphics[width=0.38\textwidth, keepaspectratio]{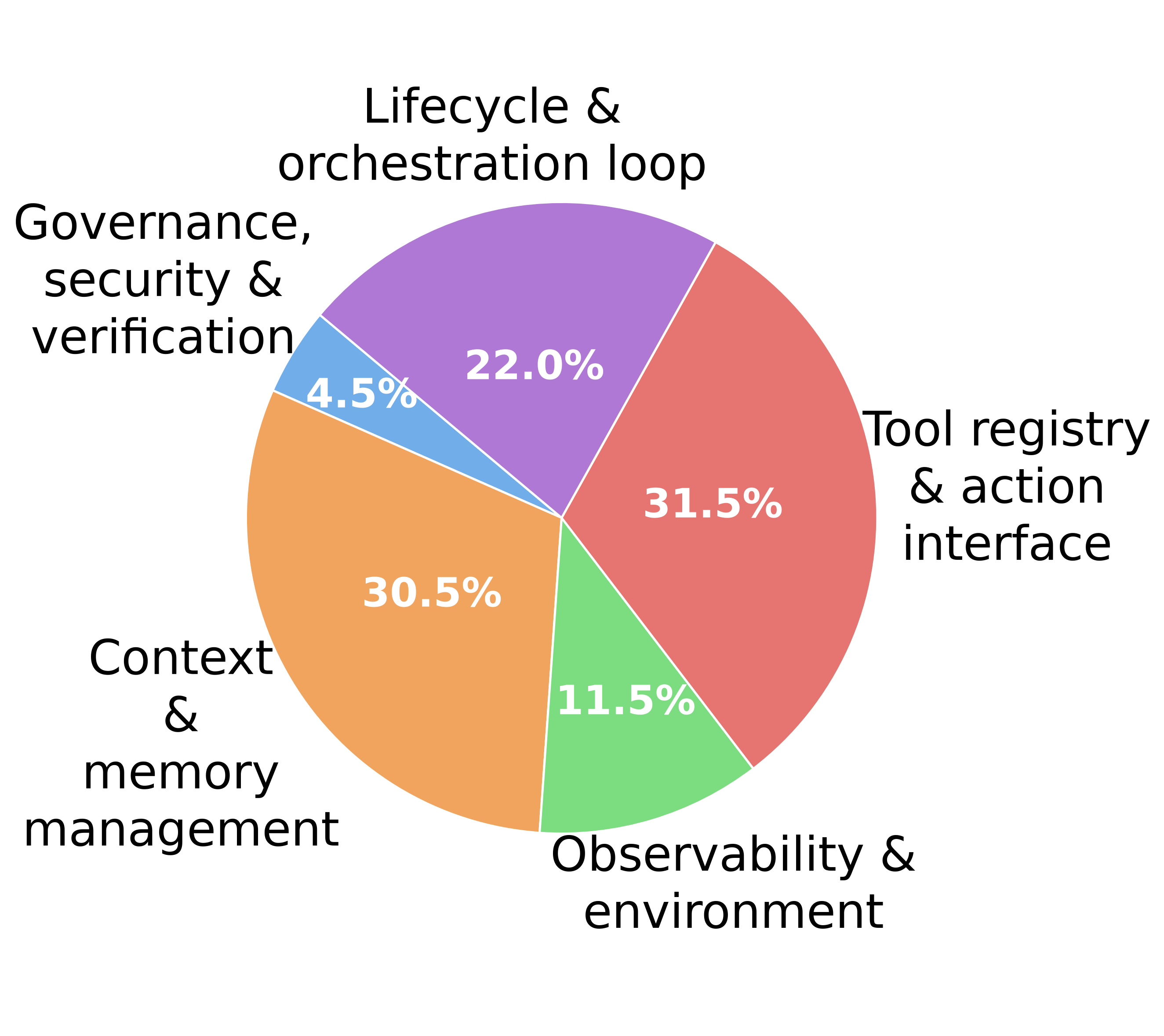}
    \captionof{figure}{Bug Distribution across Harness Components.} \label{fig:pie}
\end{wrapfigure}
The evaluation above demonstrates the effectiveness of \sysname{} in reproducing agent harness bugs. Building on this capability, we further apply \sysname{} to construct \ourbench{}, a live, large-scale benchmark of executable harness bugs. The current release of \ourbench{} contains 200 executable harness bugs and can be continuously and automatically extended with newly emerging harness bugs through \sysname{}.

\begin{figure}[!htbp]
    \vspace{0pt} 
    \centering
\includegraphics[width=0.8\textwidth,keepaspectratio]{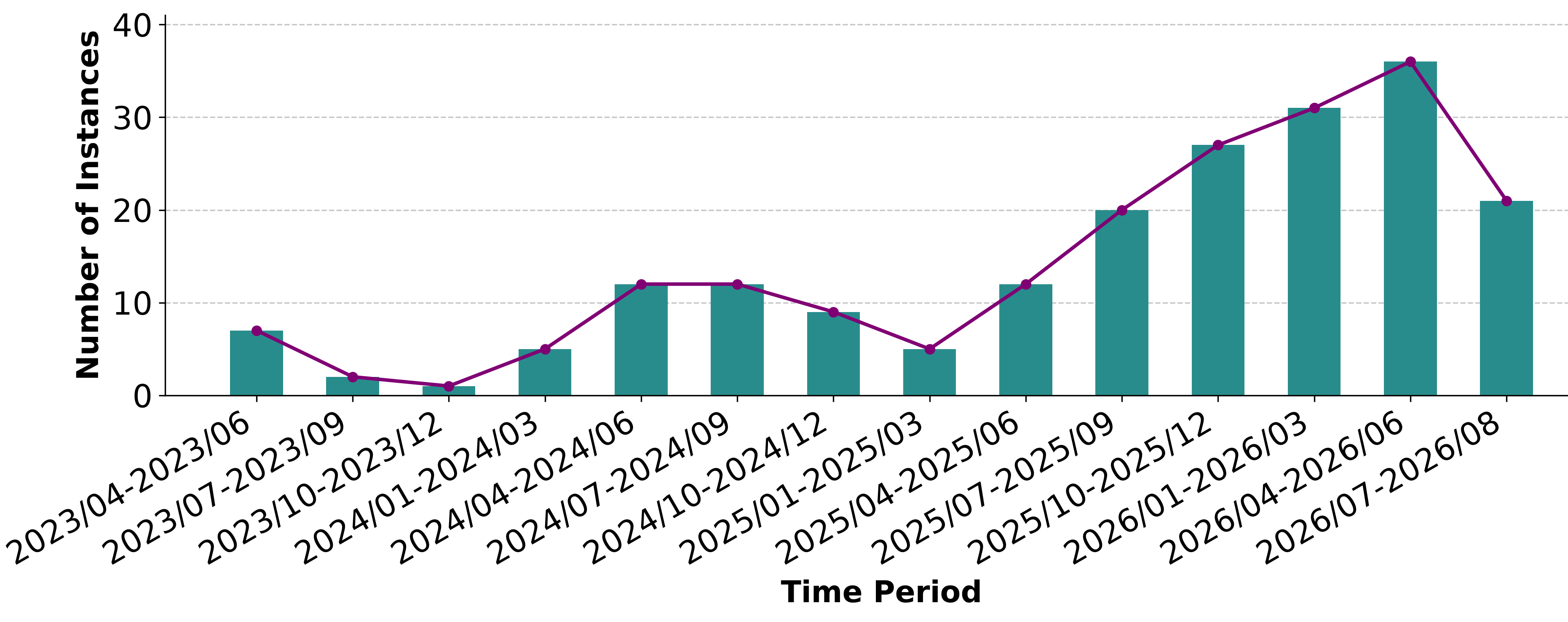}
    \captionof{figure}{Temporal Distribution of Issue Creation Times in \ourbench{}.} 
    \label{fig:bar_harness_bench}
\end{figure}

\parabf{Statistics of \ourbench{}.} Table~\ref{tab:stats-bench} summarizes the complexity of the harness bugs in \ourbench{} in terms of the scale of the corresponding code repositories and the developer-submitted gold patches. Overall, the reproduced bugs involve large codebase and non-trivial code changes, highlighting the substantial effort required to navigate the repositories, localize the faults, and implement the corresponding fixes. Figure~\ref{fig:pie} further shows the distribution of bugs in \ourbench{} across different harness components. In particular, \ourbench{} provides diverse coverage of all critical agent harness components, for example, with 31.5\% of the bugs involving tool registries and action interfaces and 30.5\% involving context and memory management. Moreover, Figure~\ref{fig:bar_harness_bench} presents the temporal distribution of the reproduced bugs, demonstrating consistent coverage across time. Enabled by \sysname{}, \ourbench{} can also incorporate recent harness bugs, including those reported as recently as August 2026. Going forward, periodically applying \sysname{} to continuously extend \ourbench{} with newly reported bugs can help mitigate concerns about contamination from model training data.

\parabf{Downstream Application I: Evaluation Benchmark.} One straightforward application of \ourbench{} is to serve as a benchmark for systematically evaluating the capabilities of existing software agents in fixing harness bugs. Table~\ref{tab:overallfix} presents the performance of three widely used software agents, \minisweagent{}~\citep{mini-swe-agent}, \openhands{}~\citep{openhands}, and \autocoderover{}~\citep{auto-code-rover}, with \gpt{} on \ourbench{}. The ``plausibly resolved'' column refers to cases where the generated patch passes the failure-reproducing tests.
Following the previous work~\citep{neuripsRahardja25} and the common practice in program repair~\citep{DBLP:conf/issta/Xia024}, we further manually inspect whether the plausible patches are semantically equivalent to the developer-submitted gold patch, which is presented in the ``correctly resolved'' column. Overall, state-of-the-art software agents exhibit limited resolution rates on \ourbench{} (e.g., at most 9.00\%), especially compared with their resolution rates reported on general software bugs, e.g., 40.67\% on SWE-Bench Verified~\citep{swebenchverified}. These results highlight the challenges of fixing real-world harness bugs and the importance of constructing reproducible harness bug benchmarks for rigorous evaluation. Detailed analysis of the resolution rates across harness components is in Appendix~\ref{app:evaluationanalysis}.

\begin{table*}[!hbt]
  \caption{Effectiveness of Software Agents on \ourbench{}.}\label{tab:overallfix}
  \centering
  \footnotesize
  \begin{adjustbox}{width=0.8\columnwidth}
  \begin{tabular}{l|c|c|cc|c}
    \toprule
    \multirow{2}{*}{\textbf{Agent}} &   \multirow{2}{1.8cm}{\textbf{Plausibly resolved\%}} & \multirow{2}{1.8cm}{\textbf{Correctly resolved\%}} & \multicolumn{2}{c|}{\textbf{Localization \%}} & \textbf{Avg.} \\ 

    &&& \textbf{File-level} & \textbf{Function-level} & \textbf{\$Cost}\\ \midrule
    
    \minisweagent & 19.50 & 9.00 & 75.12 & 68.66 & 0.38 \\ \hline

    \openhands & 17.50 & 8.50 & 81.00 & 66.00 & 0.03\\ \hline

    \autocoderover & 6.00 & 3.50 & 37.19 & 31.66 & 0.04\\ 
    \bottomrule
  \end{tabular}
  \end{adjustbox}
\end{table*}

\parabf{Downstream Application II: Enhancement.} Given the limited capabilities of existing software agents in fixing harness bugs (as shown in Application~I), another application of \ourbench{} is to serve as a knowledge base for enhancing existing software agents on harness bug repair. To validate this potential, we split \ourbench{} into training and test sets. To avoid overfitting and data leakage, we strictly ensure that harness bugs from the same agent repository do not appear in both the training and test sets. As a result, the training set contains 121 harness bugs, while the remaining 79 harness bugs form the test set. Specifically, we implement a basic skill distillation pipeline that iteratively generates and refines a textual skill (\texttt{SKILL.md}) over the training instances. We then apply the distilled skill to an existing software agent, \minisweagent{}, which achieves the best effectiveness in our evaluation in Application~I. The detailed skill distillation design is in Appendix~\ref{app:skill-protocol}. As shown in Table~\ref{tab:skillresult}, the distilled skill improves the correct resolution rate of the off-the-shelf software agent by 6.32\%, while achieving 31.81\% function-level localization accuracy. Appendix~\ref{app:skill-example} presents an example illustrating how the distilled skill enables \minisweagent{} to fix a harness bug that it cannot resolve without the distilled skill. 
Overall, given the continuously growing nature of \ourbench{}, it can serve as a scalable data foundation for continuously improving software agents on harness bug repair. More broadly, as harness improvement constitutes an important building block toward recursively self-improving agents, \ourbench{} provides a foundation for advancing this capability through continuously accumulated real-world harness failures and fixes.

\begin{table*}[!hbt]
  \caption{Effectiveness of Distilled Skills.}
  \label{tab:skillresult}
  \centering
  \small
    \begin{adjustbox}{width=0.77\columnwidth}
  \begin{tabular}{l|c|c|cc}
    \toprule
    \multirow{2}{*}{\textbf{Setting}} &
    \multirow{2}{*}{\textbf{Plausibly resolved\%}} &
    \multirow{2}{*}{\textbf{Correctly resolved\%}} &
    \multicolumn{2}{c}{\textbf{Localization \%}} \\
    &&& \textbf{File-level} & \textbf{Function-level} \\
    \midrule
    W/o skill & 35.44 & 1.27 & 79.32 & 34.60 \\
    W/ skill    & 41.77 & 7.59 & 92.31 & 66.41 \\
    \bottomrule
  \end{tabular}
  \end{adjustbox}
\end{table*}

\section{Conclusion}
This work presents \sysname{}, an automated harness bug reproduction approach that continuously discovers and reproduces real-world harness bugs from open-source agentic systems. \sysname{} substantially outperforms existing bug reproduction techniques designed for general software. Building on \sysname{}, we automatically construct \ourbench{}, a live and extensible benchmark that currently contains 200 reproducible harness bugs and can be continuously expanded with newly emerging bugs. Our two downstream application experiments further demonstrate that \ourbench{} can serve as both a rigorous evaluation benchmark and a reusable knowledge base for evaluating and improving existing software agents in fixing harness bugs, thereby contributing to the broader goal of recursively self-improving agents.

\bibliographystyle{iclr2027_conference}
\bibliography{ref/references}

\clearpage
\appendix
\raggedbottom
\section{Repository Selection Rules}\label{app:repo-rules}
\begin{partbox}{teal}{Dual-stream collection}
\small
\begin{tabularx}{\linewidth}{@{}l Y@{}}
\toprule
\textbf{Stream} & \textbf{Rule} \\
\midrule
Static
  & Query \github{} for curated awesome-lists
    (\texttt{awesome-agent}, \texttt{awesome-llm}, \texttt{awesome-ai-agents}, \ldots)
    and expand hits with the keyword family. \\
Incremental
  & Ingest GHArchive hourly files, extract \texttt{repo.name}, apply the same family. \\
Merge
  & Deduplicate the two streams. \\
\bottomrule
\end{tabularx}
\par\vspace{6pt}
\begin{tabularx}{\linewidth}{@{}l Y@{}}
\toprule
\textbf{Keyword family} & \textbf{Examples} \\
\midrule
Agent identity & agent, assistant, chatbot \\
LLM / providers & LLM, OpenAI, Anthropic, model names \\
Tools and memory & tool calling, memory, RAG, vector \\
Control & planner, orchestration, multi-agent \\
Prompting & prompt, ReAct, few-shot \\
Frameworks & CrewAI, AutoGen, LangChain, MetaGPT, AutoGPT \\
\bottomrule
\end{tabularx}
\par\vspace{2pt}
{\footnotesize We seed from community-curated awesome-lists of agentic systems and expand them with a finite keyword family. The family can later be extended by mining frequent terms from repositories that survive verification.\par}
\vspace{6pt}
\begin{tabularx}{\linewidth}{@{}l Y@{}}
\toprule
\textbf{Event} & \textbf{Counts as active} \\
\midrule
\texttt{PushEvent} & code was pushed \\
\texttt{PullRequestEvent}, \texttt{IssuesEvent} & a PR or issue moved \\
\texttt{WatchEvent}, \texttt{ForkEvent} & a star or fork \\
\texttt{CreateEvent}, \texttt{ReleaseEvent} & a ref or release \\
\texttt{MemberEvent}, \texttt{PublicEvent}, \texttt{DeleteEvent} & membership, visibility, or deletion \\
\bottomrule
\end{tabularx}
\par\vspace{2pt}
{\footnotesize We ingest GHArchive hourly files. A repository is \emph{active} if it appears in recent events above (pushes, pull requests, issues, stars, forks, releases). Activity is not equated with code change: a star and a push both indicate that the repository is currently visible, which is the signal we want for emerging systems.\par}
\end{partbox}

\vspace{2pt}
\begin{partbox}{teal}{Filtering gates}
\small
\begin{tabularx}{\linewidth}{@{}l Y@{}}
\toprule
\textbf{Gate} & \textbf{Keep if} \\
\midrule
Stars & $\ge 50$ \\
README & parseable \\
In-repo tests & at least one path matches the detector below (issue-agnostic; a hard gate) \\
\bottomrule
\end{tabularx}
\par\vspace{6pt}
\end{partbox}

\Needspace{16\baselineskip}

\section{Agentic System specification} \label{app:agentic-spec}
This is the document given to the repository judge.

\begin{specbox}{blue}
\noindent\textbf{Definition.}
An \emph{Agentic System} is a codebase that implements or contains an LLM-driven agent. Its operational logic is driven by a large language model and exhibits some of autonomous planning, memory management, environmental perception, and tool use. If a repository satisfies some of the components below, it should be considered an Agentic System.

\vspace{5pt}
\noindent\textbf{Core components.}
\par\vspace{3pt}
\begin{tabularx}{\linewidth}{@{}l Y@{}}
\toprule
\textbf{Component} & \textbf{What it covers} \\
\midrule
LLM ``brain'' & task decomposition, scheduling, conversational or action memory \\
Perception & user input, events, files, sensors, or other environment signals \\
Action / tooling & search, shell, databases, third-party APIs, plugins \\
Orchestration & single-agent loop or multi-agent collaboration \\
Provider integration & SDKs, API keys, or model configuration \\
Runtime artifacts & optional: Docker, prompts, tests, example scenarios \\
\bottomrule
\end{tabularx}

\vspace{5pt}
\noindent\textbf{Distinguishing features.}
\par\vspace{3pt}
\begin{tabularx}{\linewidth}{@{}l Y@{}}
\toprule
\textbf{Feature} & \textbf{Why it differs from conventional software} \\
\midrule
Nondeterminism & identical inputs may yield different outputs \\
External dependence & providers, tools, and resources change frequently \\
Cross-component failures & faults span LLM calls, memory, tools, and workflow \\
Prompt / context & prompt libraries and context length are first-class \\
Workflow orientation & loops and state checks may hang or run infinitely \\
\bottomrule
\end{tabularx}

\vspace{5pt}
\noindent\textbf{Hooks.}
A repository is likely an Agentic System if it contains:
\par\vspace{3pt}
\begin{tabularx}{\linewidth}{@{}c l Y@{}}
\toprule
\textbf{\#} & \textbf{Hook} & \textbf{Typical evidence} \\
\midrule
1 & LLM provider & \texttt{openai}, \texttt{anthropic}, or a custom wrapper \\
2 & Prompts & \texttt{prompt/}, \texttt{templates/}, \texttt{prompts/} \\
3 & Memory & vector DB, session store, history module \\
4 & Tools & \texttt{tools/}, \texttt{plugins/}, \texttt{tool\_wrappers/} \\
5 & Control loop & \texttt{planner}, \texttt{scheduler}, agent loop \\
6 & Model config & model name, token limit, API key, context length \\
7 & Language & README mentions of agent, planner, assistant loop, tool invocation \\
\bottomrule
\end{tabularx}

\vspace{5pt}
\noindent\textbf{Checklist.}
Three or more ``yes'': \emph{likely}. Five or more: \emph{highly likely}. The judge emits \texttt{owner/repo} and drops documentation collections, tutorials, paper lists, and standalone tool libraries.
\par\vspace{3pt}
\begin{tabularx}{\linewidth}{@{}c Y c@{}}
\toprule
\textbf{ID} & \textbf{Question} & \textbf{If yes} \\
\midrule
AS1 & README or code mentions agent, planner, tool invocation, memory, or LLM? & likely \\
AS2 & Depends on an LLM-provider SDK? & likely \\
AS3 & Prompt / template directory, or prompt-management code? & likely \\
AS4 & Memory / session / vector store? & likely \\
AS5 & Wrappers or calls to external tools or plugins? & likely \\
AS6 & Orchestration or an agent loop (planner / scheduler)? & strongly \\
\bottomrule
\end{tabularx}
\end{specbox}

\section{Issue Selection Rules}\label{app:issue-rules}

\begin{partbox}{orange}{Keep if}
\small
\begin{tabularx}{\linewidth}{@{}l Y@{}}
\toprule
\textbf{Criterion} & \textbf{Rule} \\
\midrule
Closure
  & \texttt{state\_reason} $=$ \texttt{completed};
    drop \texttt{not\_planned} and \texttt{null} \\
Description
  & non-empty body after stripping whitespace \\
Linked fix
  & at least one PR (local git if cloned, else the \github{} API) \\
Merge target
  & some linked PR is merged into \texttt{main} or \texttt{master};
    drop \texttt{develop} / \texttt{release} \\
Patch size
  & added $+$ deleted lines $\ge 20$ \\
In-patch tests
  & recorded with the same detector as in-repo tests, never required \\
\bottomrule
\end{tabularx}
\end{partbox}

\vspace{2pt}
\begin{partbox}{orange}{Drop if the patch is exclusively}
\small
\begin{tabularx}{\linewidth}{@{}l Y@{}}
\toprule
\textbf{Kind} & \textbf{Patterns} \\
\midrule
Documentation
  & \texttt{docs/}, README, \texttt{*.md}, \texttt{*.rst} \\
Lock / generated
  & \texttt{package-lock.json}, \texttt{yarn.lock}, \texttt{poetry.lock},
    \texttt{Pipfile.lock}, \texttt{Cargo.lock}, \texttt{go.sum},
    \texttt{*.pb.go}, \texttt{dist/}, \texttt{build/} \\
Vendored
  & \texttt{vendor/}, \texttt{third\_party/}, \texttt{node\_modules/} \\
Binary / assets
  & images, archives, PDF, audio/video,
    \texttt{.onnx}, \texttt{.pt}, \texttt{.pth}, \texttt{.h5}, \texttt{.pickle} \\
Formatting
  & $\ge 20$ changed lines of which $<5\%$ have non-whitespace content \\
\bottomrule
\end{tabularx}
\end{partbox}

\vspace{2pt}
\begin{partbox}{orange}{Stored fields}
\small
\begin{tabularx}{\linewidth}{@{}l Y@{}}
\toprule
\textbf{Field} & \textbf{Role} \\
\midrule
\texttt{repo}, timestamps & which Agentic System, when crawled \\
issue meta & number, title, url, body, labels \\
linked PR & number, merge flag, base branch \\
\texttt{base\_sha} & buggy snapshot (pre-merge mainline) \\
\texttt{head\_sha} & patched snapshot (PR tip) \\
\texttt{patch} & unified diff; may be missing \\
in-repo tests & \texttt{existing\_test\_paths} \\
in-patch tests & \texttt{test\_paths\_in\_patch}; optional \\
\texttt{ai\_judgment} & Harness Bug decision and raw response \\
\bottomrule
\end{tabularx}
\end{partbox}

\section{Harness Bug specification}
\label{app:harness-spec}

This is the document given to the issue judge.

\begin{specbox}{purple}
\noindent\textbf{Definition.}
A \emph{Harness Bug} is a user-reported problem (bug report or feature request) in an Agentic System that concerns agent-specific execution machinery: LLM-provider integration, tool invocation, memory, LLM operation, workflows, and utilities.

\vspace{5pt}
\noindent\textbf{Taxonomy.}
Six categories and twenty sub-categories. Category F is listed so that utility issues---failures that also arise in conventional software---can be excluded.
\par\vspace{3pt}
\begin{tabularx}{\linewidth}{@{}l >{\raggedright\arraybackslash}p{2.85cm} Y@{}}
\toprule
\textbf{ID} & \textbf{Sub-category} & \textbf{Typical failure} \\
\midrule
\multicolumn{3}{@{}l}{\textit{A.~Incompatibility with LLM providers}} \\
A.1 & Incompatible dependencies & missing or misused provider SDKs (e.g., OpenAI, LiteLLM) \\
A.2 & Unsupported models & cannot bind popular models (GPT-4, Claude, DeepSeek, \ldots) \\
A.3 & Incompatible parameters & unexpected or missing provider arguments \\
\midrule
\multicolumn{3}{@{}l}{\textit{B.~Tool-related issues}} \\
B.1 & Tool dependencies & missing libraries or binaries needed to register or run tools \\
B.2 & Tool configuration & retriever / embedder / retrieval-mode misconfiguration \\
B.3 & Tool implementation & bugs in tool or RAG logic, including helpers \\
B.4 & Misused interfaces & bad arguments, serialization, or LLM--tool binding \\
\midrule
\multicolumn{3}{@{}l}{\textit{C.~Memory-related issues}} \\
C.1 & Initialization & DB / workspace reset failures, inconsistent state \\
C.2 & Content errors & bad message attributes, serialization, non-primitive types \\
C.3 & Memory dependencies & broken internal or external memory-stack modules \\
\midrule
\multicolumn{3}{@{}l}{\textit{D.~LLM operation issues}} \\
D.1 & Model access & wrong binding or missing credentials \\
D.2 & Token usage & max tokens, pricing, or token-accounting failures \\
D.3 & Output handlers & empty, malformed, or refusal responses mishandled \\
D.4 & Model dependencies & missing tokenization or provider-client libraries \\
D.5 & Context length & overflow or incorrect length accounting \\
D.6 & Prompts & missing, stale, or poorly managed prompts \\
\midrule
\multicolumn{3}{@{}l}{\textit{E.~Workflow issues}} \\
E.1 & Scheduling / loops & hangs, infinite loops, skipped steps \\
\midrule
\multicolumn{3}{@{}l}{\textit{F.~Utility issues}} \\
F.1 & Implementation & UI, Docker, logging, unrelated imports \\
F.2 & Dependencies & non-agent libraries or internal circular imports \\
F.3 & Configuration & I/O paths, encoding, IPs, telemetry \\
\bottomrule
\end{tabularx}

\vspace{5pt}
\noindent\textbf{Checklist.}
Accepted if closed with a developer patch, at least one row is yes, and it is \emph{not} a utility-only issue.
\par\vspace{3pt}
\begin{tabularx}{\linewidth}{@{}c Y c@{}}
\toprule
\textbf{ID} & \textbf{Question} & \textbf{If yes} \\
\midrule
HB1 & Mentions an LLM provider, model name, SDK, or API key? & include \\
HB2 & Mentions prompt content, templates, or prompt management? & include \\
HB3 & Reports memory symptoms (missing history, corrupt storage, init failures)? & include \\
HB4 & Involves tool invocation, parameters, configuration, or implementation? & include \\
HB5 & Describes workflow anomalies (hangs, loops, repeated actions)? & include \\
HB6 & Patch changes LLM calls, memory, tool wrappers, prompts, or orchestration? & include \\
\bottomrule
\end{tabularx}
\end{specbox}

\section{One-to-one issue--PR--commit matching}~\label{app:one-one-match}
Issues, PRs, and commits are many-to-many and would otherwise duplicate the same fix. For each PR $p$ we collapse the set of $k$ commits into a squash commit $c_p^{s}$, build a bipartite graph $G=(\mathcal{P},\mathcal{C}^{s},E)$, and keep only one-to-one edges
\begin{equation}
\mathcal{C}(p)
=
\bigcup_{j=1}^{k}\{c_j\}
\;\longrightarrow\;
c_p^s
\qquad
E^{\star}
=
\bigl\{
(p,c^{s})\in E
\;\big|\;
\deg_{G}(p)=\deg_{G}(c^{s})=1
\bigr\}.
\end{equation}
With issue--PR incidence $E_{IP}$, each instance is $(i,p,c^{s})\in\mathcal{D}$ where $(p,c^{s})\in E^{\star}$ and $(i,p)\in E_{IP}$. The squash commit is the patched snapshot; the PR base commit is the buggy snapshot.

\section{Example of Joint Optimization }
\label{app:cofix-case}
AgentScope issue \#1297 (\url{https://github.com/agentscope-ai/agentscope/issues/1297}) reports that a disconnected Studio hook raises a \texttt{ConnectionError} after retrying and crashes the agent. Its developer patch logs the failure and returns, but it contains no IPT. This instance exposes why independent artifact refinement can stall when both the import path and the test oracle are misaligned.

\begin{partbox}{purple}{Observed trace for AgentScope issue \#1297}
\small
\begin{tabularx}{\linewidth}{@{}l Y@{}}
\toprule
\textbf{Point in pipeline} & \textbf{Observed evidence} \\
\midrule
Before co-fix & All 25 attempts across five epochs to optimizing the test still remained \texttt{f2f} (i.e., the test fails on both buggy and patched commits). In the final attempt, the patched run still loaded the hook from \texttt{site-packages}, while the test patched a logger location that did not correspond to the implementation's observable behavior. \\
Full-generation co-fix & With no IPT to restore, one response changed the container to an editable install with \texttt{/app/src} first on \texttt{PYTHONPATH}, and rewrote the test to simulate a failed \texttt{requests.post} call and assert that the real agent reply completes after the expected retries. \\
Re-verification & The same command returned exit code 1 on the buggy snapshot and returned exit code 0 after applying the patch, yielding the first \texttt{f2p} (i.e., failed on the buggy commit and passing on the patched commit) outcome for the instance. \\
\bottomrule
\end{tabularx}
\end{partbox}

This trace demonstrates the role of full-generation co-fix after test-only rounds have failed: one final diagnosis can reconcile the container's source path and the test's behavioral oracle. It is a representative execution trace, not a component-wise causal ablation; without evaluating the two crossed combinations, we do not claim that each individual edit is independently necessary.

\section{Failure-mode analysis}
\label{app:failure-modes}
Figure~\ref{table:failure-modes} summarizes the major failure modes observed in unsuccessful reproduction runs of \sysname{}. Moreover, some failure distributions vary across backbone models, suggesting that some failure modes are model-dependent. The results also indicate that further improvements may require targeted enhancements at different stages of the reproduction pipeline, such as more robust dependency resolution, environment construction, and execution-feedback collection. In particular, future efforts could prioritize improving dependency installation and Docker image construction, as well as preserving diagnostic logs when container disk exhaustion occurs.

\begin{table}[t]
\centering
\small
\setlength{\tabcolsep}{4pt}
\caption{Failure modes of unsuccessful reproduction runs. }
\label{table:failure-modes}
\begin{tabular}{lrrr}
\toprule
\textbf{Failure mode} & \textbf{\deepseek{}} & \textbf{\gpt{}} & \textbf{\kimi{}} \\
\midrule
\multicolumn{4}{l}{\textit{\cellcolor{gray!25}Single cause: environment}} \\
Dependency installation & 44 & 33 & 19 \\
Docker build configuration & 19 & 15 & 40 \\
Missing imports or incompatible APIs & 15 & 25 & 7 \\
Python packaging metadata & 15 & 27 & 4 \\
Native build or compiler & 12 & 28 & 5 \\
Container disk exhaustion & 12 & 0 & 40 \\
Missing verifier or Docker logs & 11 & 5 & 56 \\
\midrule
\multicolumn{4}{l}{\textit{\cellcolor{gray!25}Single cause: test}} \\
Oracle or exercised behavior & 17 & 30 & 4 \\
Test command crashed or timed out & 4 & 9 & 1 \\
\midrule
\multicolumn{4}{l}{\textit{\cellcolor{gray!25}Coupled cause}} \\
Environment and test together & 13 & 8 & 3 \\
\midrule
Total unsuccessful runs & 162 & 180 & 179 \\
\bottomrule
\end{tabular}
\end{table}

\section{Stability Metric }\label{app:stability}
To check stability, we independently classify each retained issue $m{=}5$ times. Let $A_i$ be the set of issues labeled \emph{Yes} in run $i$, and let $p_j$ be the fraction of \emph{Yes} answers for issue $j$. We report set agreement and normalised entropy
\begin{equation}
S_{\mathrm{set}}
=
\frac{\bigl|\bigcap_{i=1}^{m} A_i\bigr|}{\bigl|\bigcup_{i=1}^{m} A_i\bigr|}
\qquad
S_{\mathrm{entropy}}
=
1-\frac{1}{n}\sum_{j=1}^{n} H(p_j),
\label{eq:stability}
\end{equation}
where $H(p)=-p\log_2 p-(1-p)\log_2(1-p)$ is binary entropy. Both scores are $1$ when every run returns the same label for every issue.

\section{Skill Distillation Process}\label{app:skill-protocol}
This section presents the concrete process of skill distillation, which basically follow the common pipelines design in previous work~\citep{zhao2024expel,wang2024agentworkflow}. The skill $\pi$ is a human-readable \texttt{SKILL.md} document paired with a set of lightweight pattern scanners. It is neither a fine-tuned model but a plain-text instruction that the agent reads as part of its prompt. The pipeline  refines $\pi$ iteratively over the training set through the following steps.

\begin{enumerate}
    \item \textbf{Initialization.} The pipeline begins with an empty skill, 
    and the agent attempts each training issue $i \in \mathcal{I}_{\mathrm{tr}}$ 
    with no guidance beyond its default prompt.

    \item \textbf{Failure diagnosis.} When the agent fails on a training issue, 
    the pipeline diagnoses whether the failure matches a recognizable bug shape. For example, an overly narrow string predicate, an unhandled empty payload at an API boundary, a missing fallback branch, or a dropped keyword argument at a call site.

    \item \textbf{Generic scanner proposal.} If a failure matches a known shape, the skill is updated by adding or tightening a generic scanner: a short instruction that tells the agent what syntactic or semantic pattern to look for and what minimal edit to apply. Crucially, proposed updates are constrained 
    to remain within the \emph{feasible set}:
    \begin{equation}
        \Pi_{\mathrm{safe}}
        = \bigl\{\pi : \pi\ \text{contains no repository name, file path, 
        issue ID, or gold-patch text}\bigr\},
        \label{eq:skill-safe}
    \end{equation}
    so that the skill cannot memorize any specific codebase. Any proposed update 
    that references the current repository is automatically rejected:
    \begin{equation}
        \pi_{t+1}
        =
        \begin{cases}
            \pi' & \text{if } \pi' \in \Pi_{\mathrm{safe}}, \\
            \pi_t & \text{otherwise.}
        \end{cases}
        \label{eq:skill-update}
    \end{equation}

    \item \textbf{Freezing.} After one pass through all training issues, the skill $\pi_T$ is frozen. No further updates are made before evaluation on the held-out test set.
\end{enumerate}

\section{Detailed analysis of fixing effectiveness}\label{app:evaluationanalysis}
\begin{figure*}[!hbt]
    \centering
    \begin{minipage}[!hbt]{0.5\textwidth}
        \centering
        \vspace{0pt}%
        \includegraphics[width=0.95\linewidth]{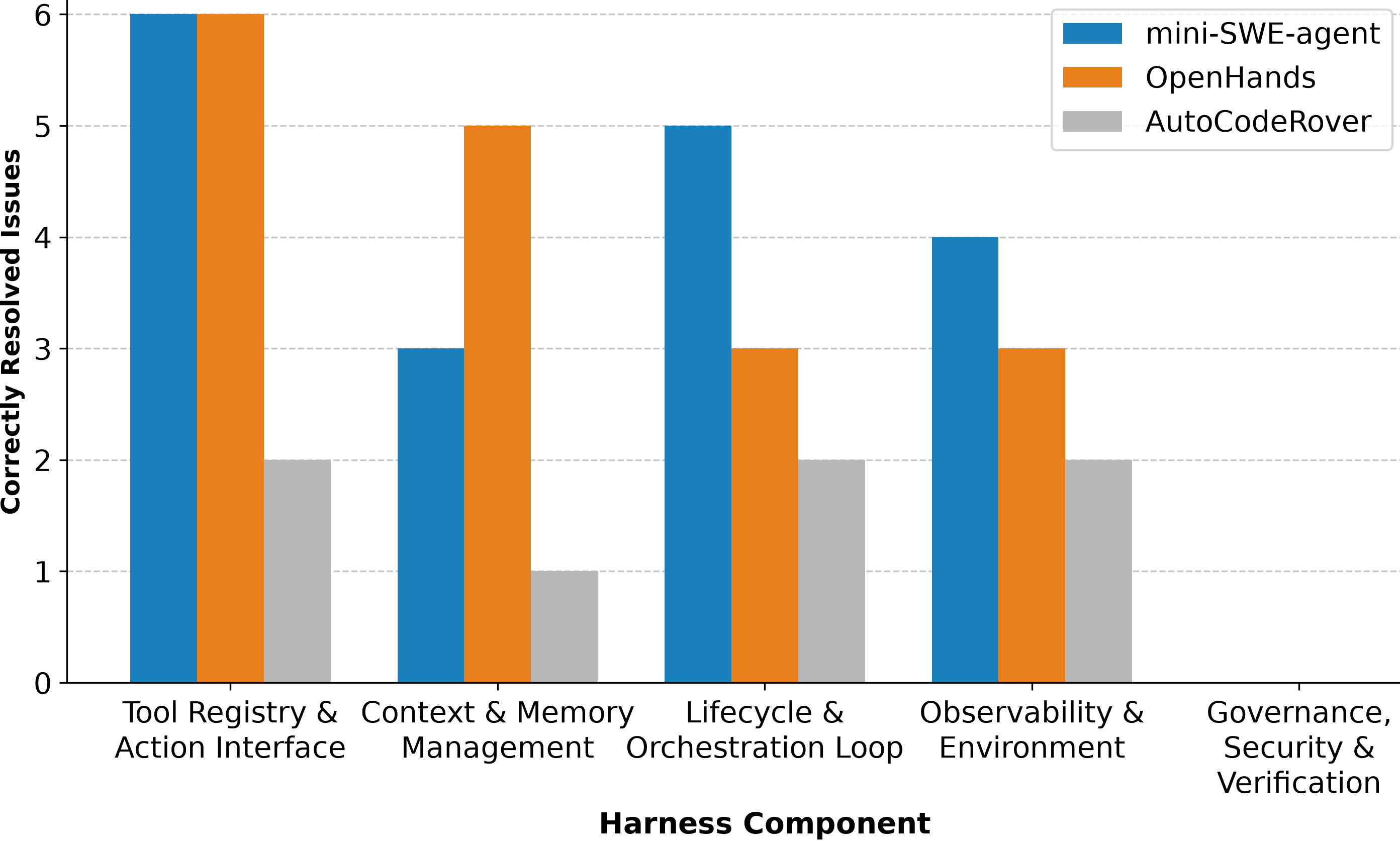}
        \vspace{0.3em}
        \captionof{figure}{Effectiveness of Agents on \ourbench{}.}\label{fig:bar_resolution}
    \end{minipage}%
    \hspace{15pt}
    \begin{minipage}[!hbt]{0.45\textwidth}
        \centering
        \vspace{0pt}%
        \includegraphics[width=0.90\linewidth]{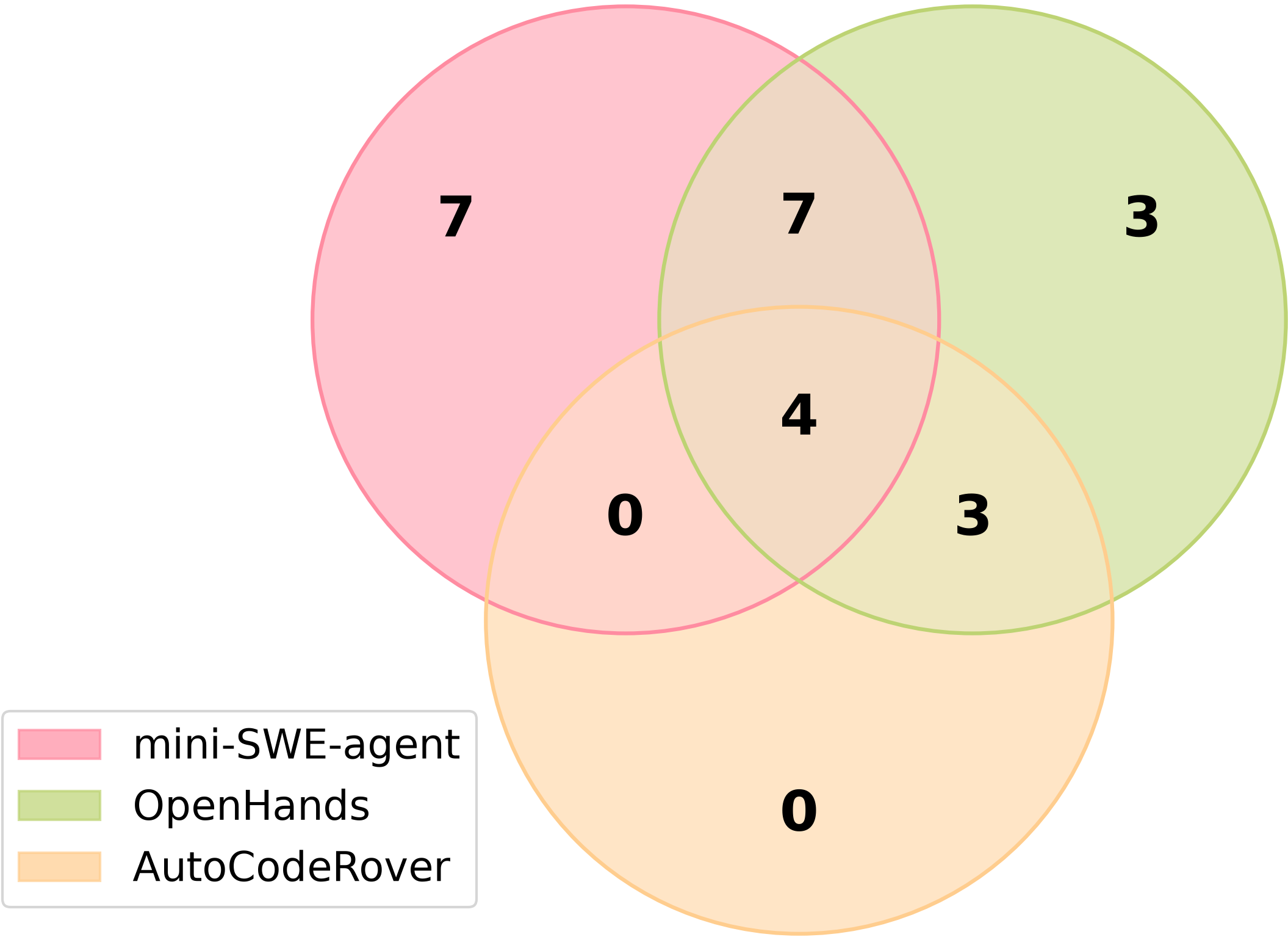}
        \vspace{0.3em}
        \captionof{figure}{Venn Diagram of Resolved Bugs.}\label{fig:venn_resolved}
    \end{minipage}
\end{figure*}

Figure~\ref{fig:bar_resolution} compares the fixing rate of studied software agents across the bugs of different harness components. Overall, tool related bugs are the one with highest fixing rate for all studied agents. Figure\ref{fig:venn_resolved} shows the unique and overlap of correctly resolved issues across the studied software agents on \ourbench{}. All the studied agents collectively resolved a total of 24 unique issues, with \minisweagent{} and \openhands{} demonstrating the highest overall efficacy by participating in 18 and 17 total resolutions respectively. \minisweagent{} achieved the strongest independent problem-solving capacity with 7 total resolutions.

\section{Case: harness-sdk \#362}
\label{app:skill-example}

This is one of the five new held-out fail-to-pass successes. C5 was not specialized to this repository.

\begin{caseouter}
\begin{caseissue}
\texttt{strands-agents/harness-sdk\#362} reports that \texttt{system\_prompt} is not passed into \texttt{structured\_output}.
\texttt{Agent.structured\_output\_async} already stores \texttt{self.system\_prompt}, then calls the model with only the schema and the messages, so a fail-to-pass spy still sees \texttt{system\_prompt=None}.
\end{caseissue}

\begin{casefail}
\casefile{src/strands/agent/agent.py $\cdot$ structured\_output\_async()}
{\ttfamily\raggedright\small
events = self.model.structured\_output(output\_model, self.messages)
\par}
\vspace{3pt}
Function-level localization misses this call. Typical attempts edit Bedrock caching, rewrite a formatter, or emit no patch.
\end{casefail}

\begin{casegen}
\casefile{src/strands/agent/agent.py:460 $\cdot$ C5, plumb a missing keyword}
\difflinem{events = self.model.structured\_output(output\_model, self.messages)}
\difflinep{events = self.model.structured\_output(}
\difflinep{\phantom{--}output\_model, self.messages, system\_prompt=self.system\_prompt)}
\end{casegen}

\begin{casegold}
\casefile{harness-sdk\#466 $\cdot$ structured\_output\_async()}
\difflinem{events = self.model.structured\_output(output\_model, self.messages)}
\difflinep{events = self.model.structured\_output(}
\difflinep{\phantom{--}output\_model, self.messages, system\_prompt=self.system\_prompt)}
\end{casegold}
\end{caseouter}

The generated call-site edit is identical to gold. The original fail-to-pass test turns green, the AI judge marks the patch \emph{correct}, and function-level localization flips from miss to hit. The other four new successes have the same chain: capability $\rightarrow$ blamed site $\rightarrow$ one-line legal edit $\rightarrow$ F2P that matches gold.

\end{document}